\documentclass[aps,prl,reprint,superscriptaddress]{revtex4-1}

\usepackage{graphicx}% Include figure files
\usepackage{dcolumn}% Align table columns on decimal point
\usepackage{bm}% bold math

\begin{document}

\preprint{APS/123-QED}

\title{Strong Correlation Effects of A-site Ordered Perovskite CaCu$_3$Ti$_4$O$_{12}$ revealed by Angle-Resolved Photoemission Spectroscopy} % Force line breaks with \\

\author{H. J. Im}
\email{hojun@cc.hirosaki-u.ac.jp}
\affiliation{Department of Advanced Physics, Hirosaki University, Hirosaki 036-8224, Japan}


\author{M. Tsunekawa}
\affiliation{Faculty of Education, Shiga University, Shiga 522-8522, Japan}

\author{T. Sakurada}
\affiliation{Department of Advanced Physics, Hirosaki University, Hirosaki 036-8224, Japan}

\author{M. Iwataki}
\affiliation{Department of Advanced Physics, Hirosaki University, Hirosaki 036-8224, Japan}

\author{K. Kawata}
\affiliation{Department of Advanced Physics, Hirosaki University, Hirosaki 036-8224, Japan}

\author{T. Watanabe}
\affiliation{Department of Advanced Physics, Hirosaki University, Hirosaki 036-8224, Japan}

\author{K. Takegahara}
\affiliation{Department of Advanced Physics, Hirosaki University, Hirosaki 036-8224, Japan}

\author{H. Miyazaki}
\affiliation{Department of Environmental and Materials Engineering, Nagoya Institute of Technology, Nagoya 466-8555, Japan}

\author{M. Matsunami}
\affiliation{UVSOR Facility, Institute for Molecular Science, Okazaki 444-8585, Japan.}
\affiliation{School of Physical Sciences, The Graduate University for Advanced Studies, Okazaki 444-8585, Japan.}

\author{T. Hajiri}
\affiliation{UVSOR Facility, Institute for Molecular Science, Okazaki 444-8585, Japan.}
\affiliation{Graduate School of Engineering, Nagoya University, Nagoya 464-8603, Japan}

\author{S. Kimura}
\affiliation{UVSOR Facility, Institute for Molecular Science, Okazaki 444-8585, Japan.}
\affiliation{School of Physical Sciences, The Graduate University for Advanced Studies, Okazaki 444-8585, Japan.}

\date{\today}% It is always \today, today,

\begin{abstract}
We report angle-resolved photoemission spectroscopy (ARPES) results of A-site ordered perovskite CaCu$_3$Ti$_4$O$_{12}$. We have observed the clear band dispersions, which are shifted to the higher energy by 1.7 eV and show the band narrowing around 2 eV in comparison with the local density approximation calculations. In addition, the high energy multiplet structures of Cu 3$d^8$ final-states have been found around 8 - 13 eV. These results reveal that CaCu$_3$Ti$_4$O$_{12}$ is a Mott-type insulator caused by the strong correlation effects of the Cu 3$d$ electrons well hybridized with O 2$p$ states. Unexpectedly, there exist a very small spectral weight at the Fermi level in the insulator phase, indicating the existence of isolated metallic states.

%\begin{description}
%\end{description}
\end{abstract}

\maketitle

%\section{Introduction}%%%INTRODUCTION%%%

A-site ordered perovskite CaCu$_3$Ti$_4$O$_{12}$ (CCTO) has generated considerable interest due to the extremely high dielectric constant ($\epsilon$) as high as 10$^4$-10$^5$ over a wide range of temperature from 100 to 600 K, which holds a promise for high performance capacitor \cite{Subr00}.
Prior to applications, there have been many studies to identify the intrinsic mechanism of the high $\epsilon$.
Although the consistent conclusion of the origin has not been established yet, it has been widely accepted that the high $\epsilon$ would come from defects and/or disorder structures, e.g. a relaxor like dipole fluctuation in nanosize domain \cite{Home01}, an internal barrier layer capacitance \cite{Chun04}, and the nanoscale disorder of Ca and Cu-site \cite{Zhu07}.
Generally, the origin of the high $\epsilon$ of CCTO has been considered to be different from that of conventional ferroelectric materials, because of the absence of structural transition accompanying with the abrupt change of $\epsilon$ around 100 K \cite{Rami00,Litv03}.
Therefore, in order to understand these intrigue physical properties, the electronic structures should be clarified experimentally.
In particular, relations between the electronic structure and the strong correlation effects are central issues \cite{Imad98, Kotl06, He02}.  
For example, the high $\epsilon$ and an insulator phase of CCTO can not be explained by theoretical calculations within the local density approximation (LDA), which are not considered to properly treat strong correlation between electrons.
The strong correlation effects can be also expected from the crystal structure of CCTO, which contains the CuO$_4$ plane units in similar to the CuO$_2$ plane of the high-$T_c$ cuprates as shown in Fig. 1 (b) \cite{Long09}.
Recently, it has been reported that a family compound, CaCu$_3$Ru$_4$O$_{12}$, shows the heavy fermion behavior and the non-Fermi liquid, supporting the importance of the strongly correlation effects \cite{Koba04}.
Hence, it has been believed that CCTO would be a Mott-type insulator, even though the experimental band dispersions have never been observed.

Here, we first report the clear observation of band dispersions of CCTO by the angle-resolved photoemission spectroscopy (ARPES) measurements.
%\section{Experiments} %%%experiments%%%
ARPES experiments were performed at the beamline BL5U of UVSOR, using photon energies ($h\nu$) from 30 to 93 eV.
Measurements were carried out at room temperature (T = 300 K) in a vacuum better than $2 \times 10^{-8}$ Pa.
Total energy resolution ($\Delta E$) and momentum resolution ($\Delta k$) are about 165 meV and 0.02 $\textrm{\AA}^{-1}$ at $h\nu$= 90 eV, respectively.
The surface of CCTO prepared by \textit{in situ} cleaving has been very stable during a typical measurement period of 12 hours, showing no sign of the progress of degradation .
In addition, the detailed electronic structure near $E_F$ has been investigated by the low-energy anlge-integrated photoemission spectroscopy (AIPES) measurements at the beamline BL7U, using $h\nu$ = 7 eV with $\Delta E$ $\sim$ 15 meV.

%\section{Results and Discussion} %%%RESULTS%%%

% >>> Figure 1 <<<

\begin{figure}
\includegraphics[width=85mm]{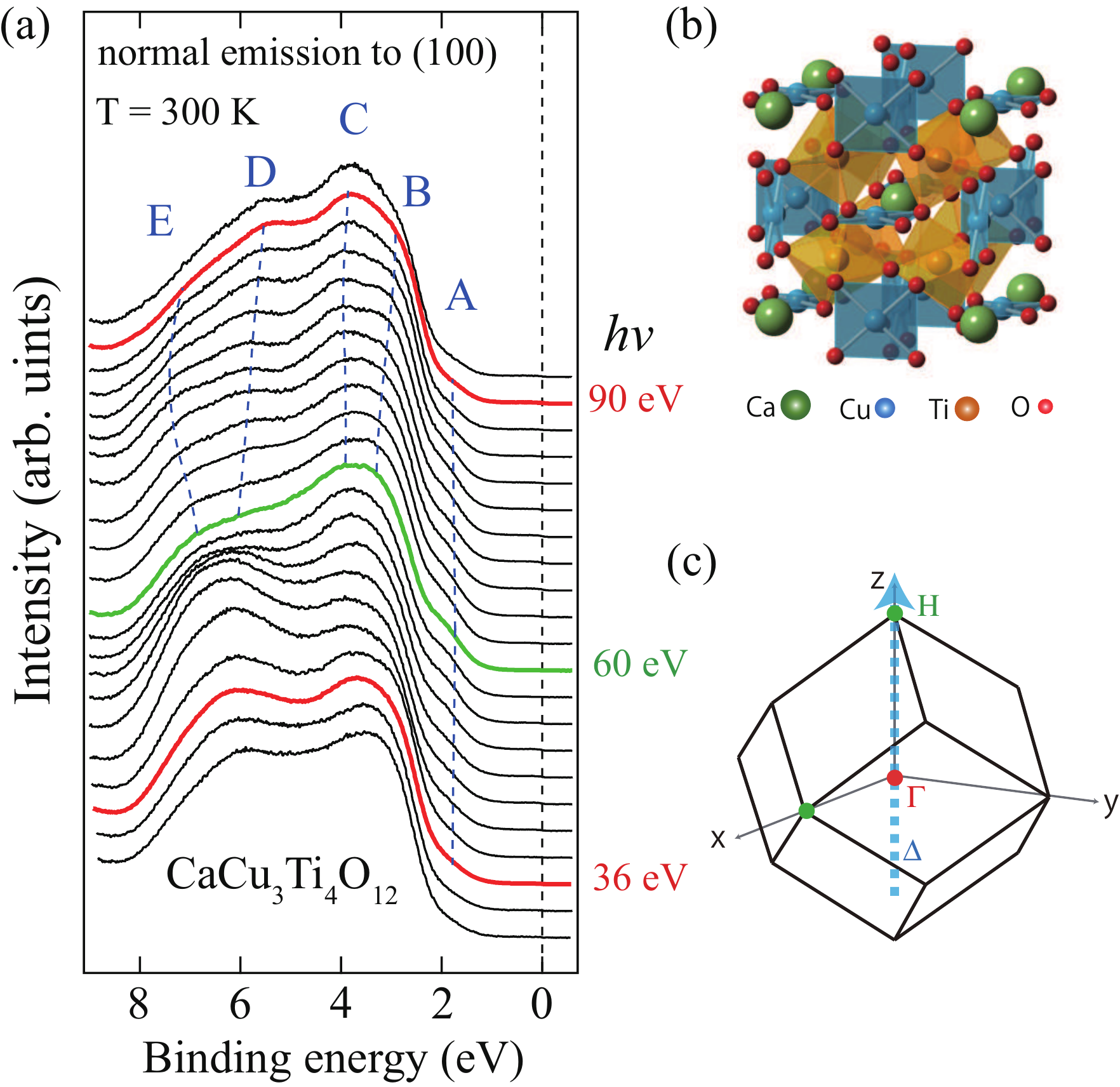}% Here is how to import EPS art
\caption{\label{fig:figure1} (color online) (a) The EDCs along $\Gamma$H ($\Delta$) - direction in the normal emission to the (100) plane, observed by using the photon energies from 30 to 93 eV at intervals of 3 eV. The blue dashed lines are guides for the eyes. (b) Body-centered cubic structure \cite{Long09} and (c) Brillouin zone of A-site ordered peroviskite CaCu$_3$Ti$_4$O$_{12}$.}
\end{figure}

Figure 1 (a) shows the energy distribution curves (EDCs) of CCTO in the valence-band region.
The spectra have been obtained at room temperature (T = 300 K) with changing $h\nu$ from 30 to 93 eV  in the normal emission to the (100) plane.
With increasing $h\nu$, ARPES spectra trace the blue arrow along $\Delta$-direction in the Brillouin zone of CCTO as depicted in Fig. 1 (c).
For the sake of convenience, the valence bands are divided into three regions, 0 - 2.5 eV, 2.5 - 5 eV, and 5 - 8 eV.
The bands in the regions of 2.5 - 5 eV and 5 - 8 eV relatively highly disperse with intense features, while the bands in the region of 0 - 2.5 eV are not well distinguished due to weak intensity and broad band width.
In the region of 0 - 2.5 eV, we observe the intensity variation of the small shoulder as a function of $h\nu$ (A).
The shoulder of A becomes prominent with changing $h\nu$ from 30 to 60 eV, and then its intensity become smaller from $h\nu $ = 60 to 90 eV.
This indicates that $h\nu$ = 60 eV can be a high symmetry point, even though we should be careful of the variation of the photoionization cross section ($\sigma$) as $h\nu$ changes.

In the region of 2.5 - 5 eV, there are two types of bands designated by the letters B and C.
The band B disperses from 3 to 2.5 eV with the top at $h\nu$ = 90 eV and the bottom at $h\nu$ = 60 eV, while the band C shows very small dispersion around 3.8 eV reflecting the localized character.
In the region of 5 - 8 eV, bands well split into two types (D and E).
As $h\nu$ closes to 90 eV, the band D has the top at 5 eV  and the band E has the bottom at 7 eV.
From the above results and the analysis based on the free-electron final-state model \cite{Hufn95}, we determined the high symmetry points [Fig. 1 (a)]:
the EDCs at $h\nu$ = 36 eV and 90 eV correspond to $\Gamma$-point (red line) and that of $h\nu$ = 60 eV to H-point (green line).
Beyond the above band dispersion, there is large intensity variation around 6.5 eV as $h\nu$ closes to about 50 eV.
Such behavior has been also observed in the other transition-metal oxides, where it comes from the resonance around O 2$p$ edge and is not relevant to band dispersion \cite{Breu95}.
At the same time, in this region, it means that the spectral weight includes much weight of O 2$p$ orbital.

% >>> Figure 2 <<<

\begin{figure}
\includegraphics[width=65mm]{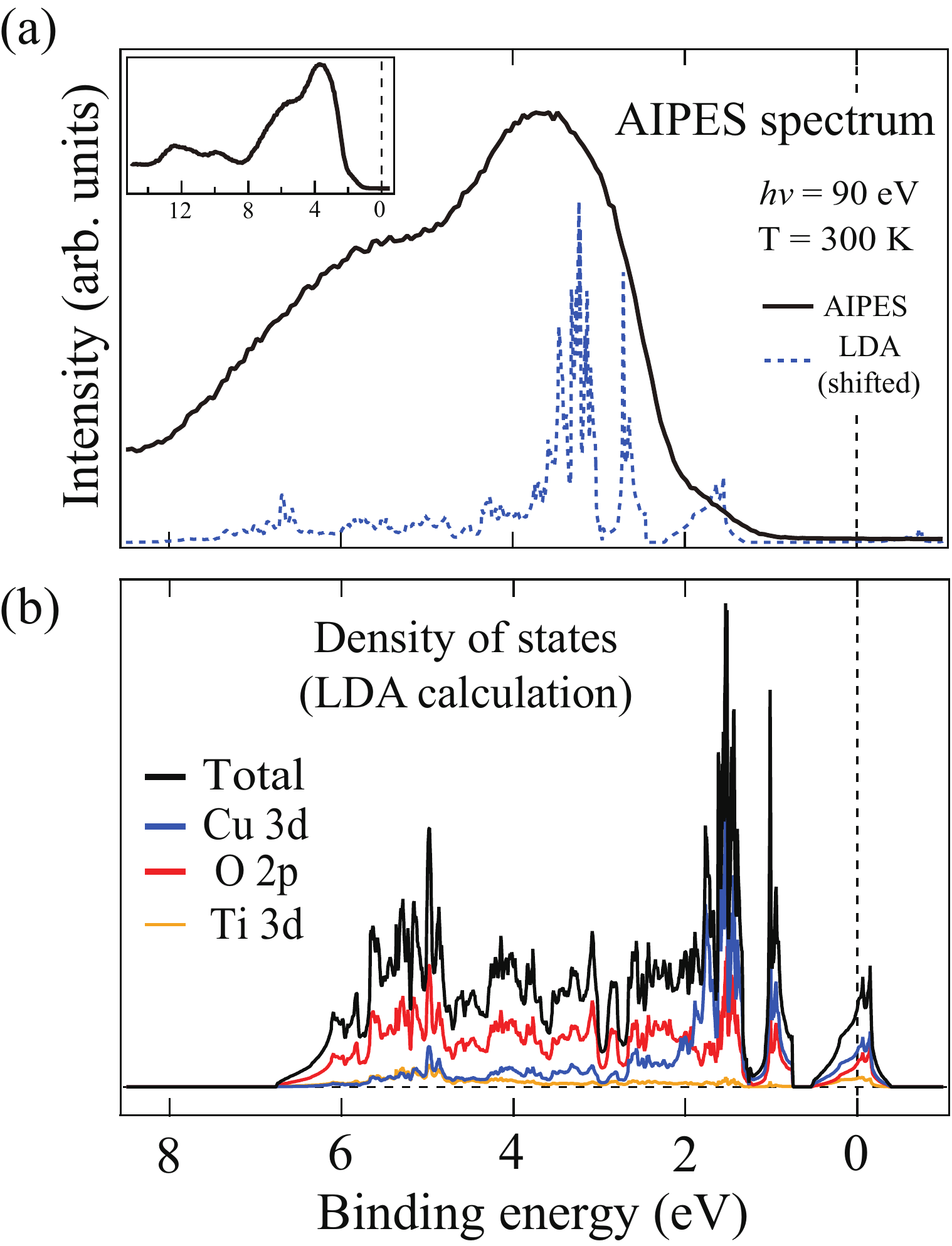}% Here is how to import EPS art
\caption{\label{fig:figure2}(color online) (a) AIPES spectrum of CCTO obtained at $h\nu$ = 90 eV and T = 300 K in valence band region. The inset shows the AIPES spectrum in wide valence band region. For comparison, the partial DOS of Cu 3$d$ is inserted below the AIPES spectrum. (b) Total DOS and the partial DOS for the valence bands of Cu 3$d$, O 2$p$, and Ti 3$d$ in the LDA calculations.}
\end{figure}

\begin{figure*}
\includegraphics[width=130mm]{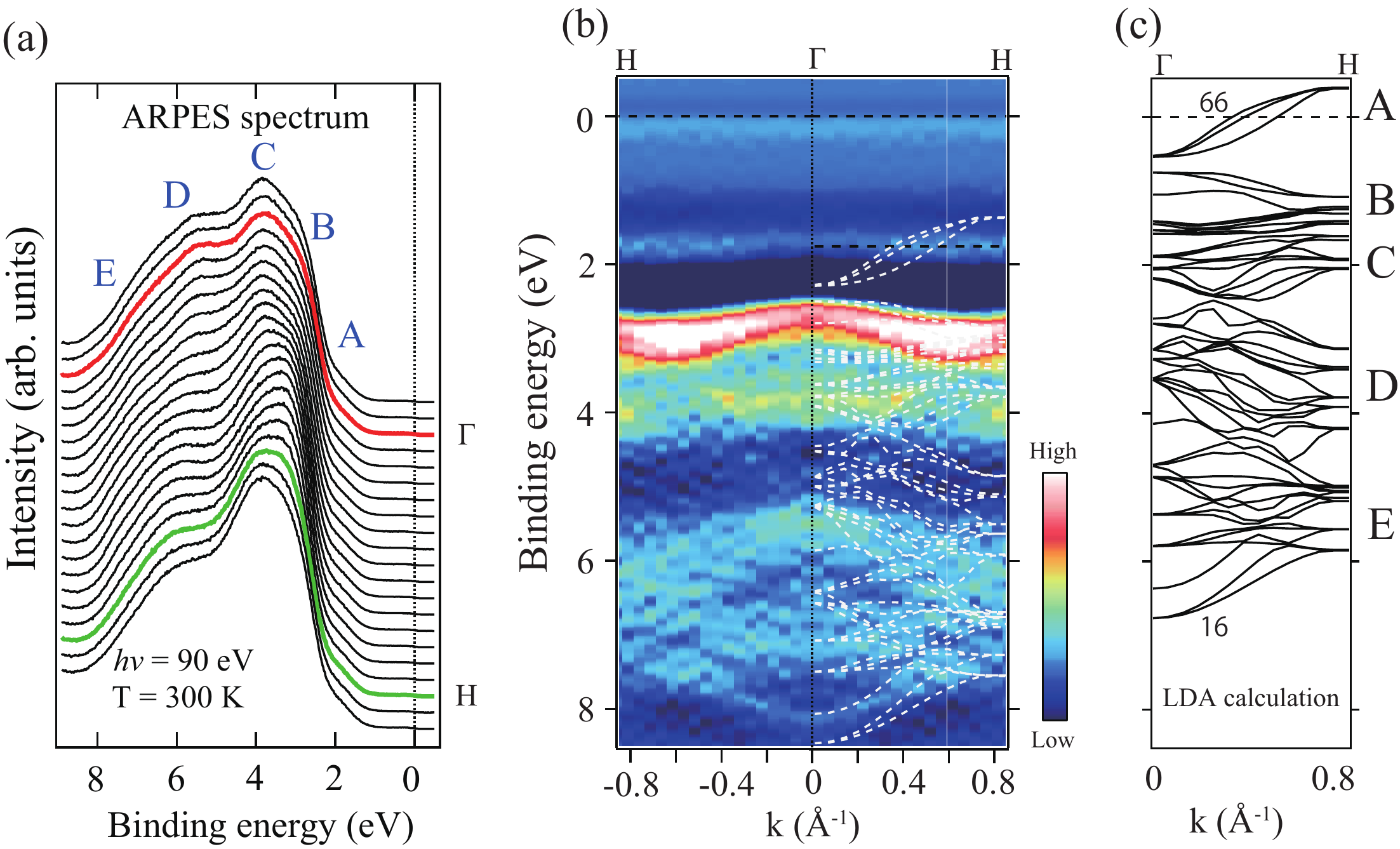}% Here is how to import EPS art
\caption{\label{fig:figure3} (color online) (a) ARPES spectra of CCTO along $\Delta$ - direction in the parallel to the cleaved surface (100) plane at $h\nu$ = 90 eV and T = 300 K. (b) ARPES image obtained from the second derivatives of the EDCs. The theoretical band dispersions (white dash line), shifted by 1.7 eV, are plotted on the ARPES image. (c) Theoretical band dispersions obtained from the LDA calculations.}
\end{figure*}

Let us compare the spectral weight obtained in the angle-integrated mode of analyzer with the density of states (DOS) in the LDA calculations.
Figure 2 (a) and 2 (b) show the AIPES spectrum at $h\nu$ = 90 eV and the partial DOS obtained from the LDA calculations, respectively.
First, we recognize that the spectral weights near $E_F$ do not seem to exist in this scale plot (actually, there are very small spectral weights, and we will discuss it later) explaining the insulating properties of the electrical resistivity experiments, while the DOS of mainly Cu 3$d$ and O 2$p$ states exits at $E_F$ in the LDA calculations.
When we compare ARPES spectra with the DOS, the variation of $\sigma$ with $h\nu$ be taken into account:  
the $\sigma$ of $d$-orbital is generally much larger than that of $p$-orbital by about ten times at $h\nu$ = 90 eV \cite{Yeh85}.
In Fig. 2 (a), we plot the partial DOS of Cu 3$d$ of the LDA calculations inside AIPES spectrum, which was shifted to higher binding energy by 1.7 eV. 
The AIPES spectrum shows good agreement with the shifted partial DOS of Cu 3$d$ in the LDA calculations.
We find that the shoulder around 2 eV with weak intensity (band A) corresponds to the DOS at $E_F$ in the LDA calculations.
There are the intense peaks in the region from 2.5 to 8 eV which correspond to the bands B, C, D, and E in Fig. 1 (a).
The bands B and C mainly consist of Cu 3$d$ states, and the band D and E mainly come from O 2$p$ states.
On the other hand, most DOS of Ti 3$d$-orbital are located in the unoccupied region [Fig. 2 (b)].
In the inset of Fig. 2 (a), we have observed two peaks around 8 - 13 eV, which do not appear in the LDA calculations.
These peaks are the intrinsic features of CCTO, because the peaks have appeared in the fresh surface just after the cleaving sample and show no degradation during the measurements.
In fact, it has been reported in CuO, family compound CCRO, etc. that such high energy peak structures come from Cu 3$d^{8}$ final-states due to multiplet effects, indicating atomic-like behaviors caused by strong correlation effects \cite{Imad98, Ghij88, Eske90, Holl12}.

% >>> Figure 3 <<<

Figure 3 (a) shows EDCs obtained in the angle-resolved mode of analyzer at $h\nu$ = 90 eV and T = 300 K.
Geometrically, the direction of the spectra is parallel to the sample surface (namely, perpendicular to the normal emission) and is also along $\Delta$-direction due to the symmetry of the cubic structure.
Hence, both EDCs in Fig. 1 (a) and in Fig. 3 (a) represent the same band dispersions.  
Certainly, we recognize that the EDCs of the $h\nu$-dependent PES from 60 to 90 eV are very similar to the ARPES spectra, indicating the same band dispersions.
At the same time, this means that our experimental results are very reliable.
Here, we can explicitly determine the band dispersion of CCTO by comparing these two kinds of photoemission spectra with the band calculations.
Figure 3 (b) is the image of the ARPES spectra obtained by the second derivatives of the EDCs.
Figure 3 (c) shows the band dispersions obtained from the LDA calculations.
And, the calculated band dispersions are superimposed on the ARPES image, shifting to the higher binding energy by 1.7 eV [Figure 3 (b)].
First, it should be noted that there are 51 bands in the valence band region as shown in Fig. 3 (c).
Unfortunately, the PES experiments cannot resolve these bands very closed each other due to the limitation of $\Delta E$ and $\Delta k$.
Therefore, the observed 5 types of band dispersions (A, B, C, D, and E) in ARPES experiments should be interpreted as a bundle of bands with the similar tendency. 
In the region of 0 - 2.5 eV, the EDCs show a little different behavior between ARPES and $h\nu$-dependent PES.
The intensity variation of the band A around 2 eV in the ARPES measurements [Fig. 3 (a)] is small compared to the $h\nu$-dependent PES measurements [Fig. 1 (a)], which may come from the different transition probability between initial state and final state in the different methods of PES measurements.
As discussed in Fig. 2 (a), we can assign the band A to three bands from 64 to 66 which cross $E_F$ and show the band dispersion of about 1 eV [Fig. 3 (c)].
Even though the exact size of band dispersion can not estimated due to the band broadening in the ARPES measurements, we find that the band width in ARPES  is very narrow compared to the LDA calculations [Fig. 3 (c)].
This indicates that the band A has the larger effective mass and is more localized than the expectation of the LDA calculation, as observed in strongly correlated $f$-electrons systems \cite{Im08}.
According to Mott-Hubbard model, the large repulsive Coulomb interaction separates the DOS into the upper Hubbard band in unoccupied region and the lower Hubbard band in occupied region \cite{Imad98}.
In the case of CCTO, the hybridized bands of Cu 3$d$ and O 2$p$ states were shifted to the higher binding energy by about 1.7 eV in comparison with the LDA calculations, showing a little different behavior compared to the simple Mott-Hubbard model.
However, this is not surprising because the Cu 3$d$ states have about 9 electrons in the occupied region and about 1 electron in the unoccupied region and are well hybridized with O 2$p$ states in spite of the strong correlation effects.
In fact, the similar band structures can be found in earlier studies of CuO and the high-T$_c$ cuprates \cite{Ghij88, Imad98}. 
In the region of 2.5 - 5 eV, the band C disperses from 3.8 eV at $\Gamma$-point (red line) to 3.5 eV at H - point (green line) showing small upturn behavior [Fig. 3 (a)].
Around 2.7 eV, there is a shoulder which corresponds to the band B.
The band B and C are assigned to bands from 60 to 63 and bands from 55 to 59, respectively. 
In the region of 5 - 8 eV, ARPES data also show the well splitted bands, D and E.
These bands consist of many bands from 16 to 50.

In summary, the tendency of band dispersions of ARPES experiments is well consistent with those of the LDA calculations, except for the band shifting to the higher binding energy by about 1.7 eV, the band narrowing around 2 eV, and the high-energy multiplet structures of Cu 3$d^8$ final-states around 8 -13 eV.
And these difference of the electric structure between ARPES experiments and LDA calculation can be assigned to the strong correlation effects as discussed above.

% >>> Figure 4 <<<

\begin{figure}
\includegraphics[width=70mm]{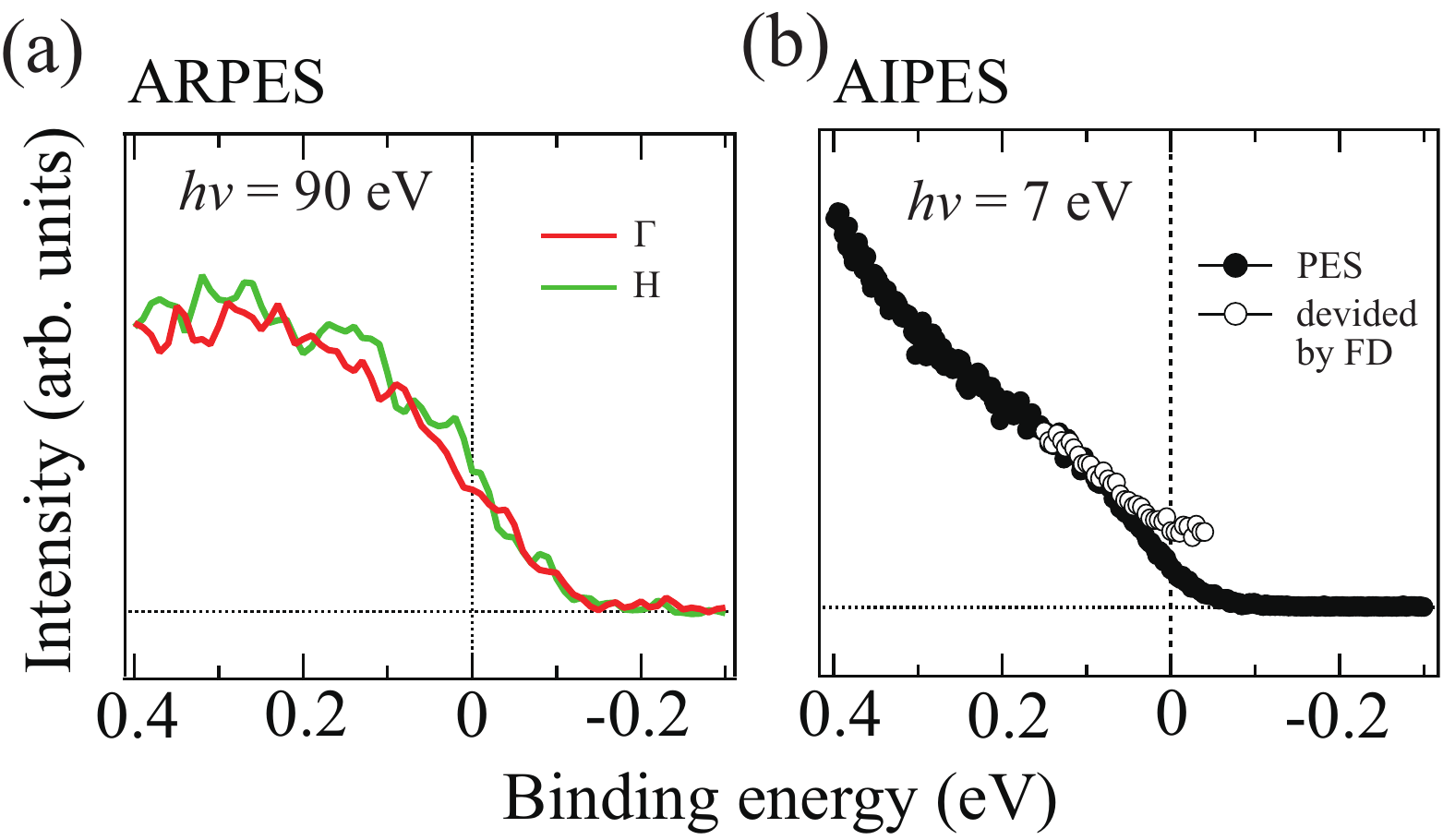}% Here is how to import EPS art
\caption{\label{fig:figure4} (color online) (a) The enlarged ARPES spectra at $\Gamma$ and H - points near $E_F$ in Fig. 3 (a). (b) The bulk sensitive AIPES spectra near $E_F$ at $h\nu$ = 7 eV. The solid circle is the AIPES sepctrum and the open circle is the spectral weight near $E_F$ obtained from dividing the AIPES spectrum by the convoluted Fermi Dirac function ($\Delta E$ $\sim$ 15 meV and T = 300 K).}
\end{figure}

Finally, let us discuss the small spectral weights at $E_F$.
Figure 4 (a) is the enlarged plot of EDCs at $\Gamma$ and H - points in Fig. 3 (a).
We find that there is a very small spectral weight at $E_F$ in spite of the insulating phase in the electrical resistivity measurements.
A possibility, Ti 3$d$ states remains at $E_F$, should be ruled out because CCTO would show the metallic properties if Ti 3$d$ states exist at $E_F$.
The another possibility is the metallic phase caused by the surface state of CCTO, because the PES spectra, obtained by using $h\nu$ = 20 - 100 eV, are surface sensitive as explained by the universal curve \cite{Seah79}.
It is well known that the mean free path in solids becomes long enough to probe the bulk properties by using the low $h\nu$ \cite{Kiss05}.  
Therefore, we have performed the AIPES measurements at $h\nu$ = 7 eV as shown in Fig. 4 (b).
The spectral weight at $E_F$ (open circle) has been recovered by dividing the AIPES spectra (solid circle) by the Fermi Dirac function convoluted by the energy resolution and the measurement temperature \cite{Greb97, Im08}.
There exist the unambiguous spectral weights at $E_F$.
For the explanation of both the PES and electrical resistivity measurements, the existence of isolated metallic states in bulk is very reasonable, which has been suggested as one of the origins of the extremely large $\epsilon$ \cite{Zhu07}.

%\section{Summary}
 %\end{document}
 
\begin{acknowledgments}
The authors are grateful to S. Nakajima for technical assistance. 
\end{acknowledgments}

%\newpage %Just because of unusual number of tables stacked at end
\bibliography{CCTOARPES_Ref}% Produces the bibliography via BibTeX.

\end{document}